\documentclass[12pt]{article}

\usepackage[utf8]{inputenc}
\usepackage[T1]{fontenc}
\usepackage{newtxtext,newtxmath} 
\usepackage{amsmath,amsfonts}
\usepackage{graphicx}
\usepackage{booktabs}
\usepackage{array}
\usepackage{longtable} 
\usepackage{verbatim}
\usepackage{url}
\usepackage[margin=1in]{geometry}
\usepackage{setspace}
\usepackage{enumitem} 
\usepackage{framed}
\usepackage{xcolor}
\definecolor{shadecolor}{RGB}{255,255,150}
\usepackage[round]{natbib}
\setcitestyle{aysep={}}

\usepackage[colorlinks=true, linkcolor=black, citecolor=black, urlcolor=black]{hyperref}

\makeatletter
\renewcommand\hyper@natlinkbreak[2]{#1}
\makeatother

\setlist[enumerate]{font=\bfseries}

\usepackage{titling}
\pretitle{\begin{center}\LARGE\bfseries}
\posttitle{\end{center}\vskip 0.5em}

\usepackage{titlesec}
\titleformat{\section}{\normalfont\Large\bfseries}{}{0pt}{}
\titleformat{\subsection}{\normalfont\large\bfseries}{}{0pt}{}
\titleformat{\subsubsection}{\normalfont\normalsize\bfseries\itshape}{}{0pt}{}
\begin{document}

\begin{center}
{\small Accepted manuscript. Published in the \textit{Journal of Responsible Innovation} (2026), \url{https://doi.org/10.1080/23299460.2026.2743406}. Please cite the published version.}\\[1em]
{\LARGE\bfseries Epistemic Trust as a Mechanism for Ethics Integration: Failure Modes and Design Principles from 70 Moral Imagination Workshops}\\[1.5em]
{\Large Benjamin Lange\textsuperscript{a,b}, Geoff Keeling\textsuperscript{c}, Kyle Pedersen\textsuperscript{d}, Carmen Heringer\textsuperscript{e}, Susan B. Rubin\textsuperscript{e,f}, Ben Zevenbergen\textsuperscript{e,g}, Amanda McCroskery\textsuperscript{h}}\\[1em]
{\small\itshape \textsuperscript{a}Faculty of Philosophy, Philosophy of Science and Religious Studies, Ludwig-Maximilians-Universit{\"a}t M{\"u}nchen, Munich, Germany; \textsuperscript{b}Munich Center for Machine Learning, Munich, Germany; \textsuperscript{c}Google, Paradigms of Intelligence Team, London, UK; \textsuperscript{d}Google Research, New York, NY, USA; \textsuperscript{e}Work done while at Google, San Francisco, CA, USA; \textsuperscript{f}The Ethics Practice; \textsuperscript{g}Stanford University, 2026 Ethics \& Technology Practitioner Fellow, Stanford, CA, USA; \textsuperscript{h}Google DeepMind, London, UK}\\[1em]
{\small Corresponding author: Benjamin Lange, \href{mailto:benjamin.lange@lmu.de}{benjamin.lange@lmu.de}}\\[2em]
\end{center}

\noindent\textbf{Abstract:} Bottom-up responsible innovation initiatives aim to empower technology development teams to engage in ethical reflection, yet such interventions frequently fail to achieve practitioner engagement. This paper proposes \emph{epistemic trust}---the degree to which practitioners regard an intervention, its facilitators, and its content as credible---as a conceptual model linking intervention design to engagement. Drawing on philosophical work on testimony and practice-based qualitative analysis of over 70 moral imagination workshops with engineering teams between 2019 and 2025, we identify five dimensions of epistemic trust (Relevance, Inclusivity, Agency, Authority, and Alignment) and present a typology of 24 failure modes that arise when these dimensions are inadequately addressed. We derive nine design principles for cultivating epistemic trust, grounded in our operationalization of moral imagination through technomoral scenarios and structured deliberation. Our findings contribute to the literature on collaborative socio-technical integration by specifying the conditions of uptake that existing frameworks leave undertheorized.\\[1em]


\vspace{2em}

\section{Introduction}

The pace of technological development exceeds the pace of its regulation \citep{Moor1985, Downes2009, Garrett2020, Umbrello2023}. Technology companies therefore frequently operate in a policy vacuum in which existing regulation fails to provide clear guidance on emerging ethical issues. In response, academics, policymakers, activists, and industry leaders have argued for greater ethical oversight in the development of new technologies via internal review processes, internal policy making, and external oversight boards \citep{Jackman2016, Shneiderman2021, Prunkl2021}. Increasingly, these calls extend beyond traditional top-down compliance controls \citep{Kaptein2008, Kaptein2011} to include bottom-up responsible innovation initiatives that aim to empower technology development teams to become proactive agents of ethical reflection and responsible innovation \citep{Owen2012, Stilgoe2013, Fiesler2021}.

Such bottom-up interventions face a central challenge: achieving perceived legitimacy among the practitioners they seek to engage. This challenge has become increasingly urgent as responsible innovation is institutionalized in research policy while public anxiety about unintended consequences persists \citep{Guston2014}. Meanwhile, research on engineering culture suggests the challenge is substantial. For example, \citet{Cech2014} demonstrates that engineering education cultivates a ``culture of disengagement'' in which concern for public welfare declines over the course of professional socialization and \citet{Lee2019} have documented the ``Rare Bird Problem'', according to which explicit ethical deliberation rarely occurs spontaneously, despite an environment that encourages open discussion of technical matters. When prompted, engineers exhibited narrow conceptions of their responsibilities and frequently deflected ethical concerns to managers or legal systems. This suggests that issuing frameworks or mandating review is not sufficient because neither produces the situated deliberation that responsible innovation requires. 

A growing body of scholarship addresses this challenge through so-called collaborative approaches to socio-technical integration \citep{Fisher2015}. These approaches aim to embed ethical and social reflection within technical practice instead of treating it as external commentary. Methods such as Socio-Technical Integration Research \citep{Fisher2013}, the Toolbox Project \citep{ORourke2013}, and Value Sensitive Design \citep{Friedman2019} have shown that structured interventions can surface ethical considerations and, in some cases, modulate research practice. However, this literature primarily describes integration methods rather than analyze \textit{why} some interventions succeed in engaging practitioners while others are dismissed as irrelevant, adversarial, or disconnected from real work.

This paper proposes that the concept of \emph{epistemic trust} helps to address this gap. Based on philosophical work on testimony and expertise \citep{Hardwig1985, Goldman1999, Origgi2004, Zagzebski2012, McCraw2015, Irzik2019} and psychological research on trust \citep{Fonagy2014, Hendriks2015}, we conceptualize epistemic trust as the degree to which practitioners regard an intervention, its facilitators, and its content as credible, relevant, and actionable. We propose that effective responsible innovation interventions must be designed to cultivate epistemic trust across multiple dimensions and that characteristic failure modes arise when these dimensions are inadequately addressed. Our argument combines conceptual development with practice-based inquiry.

We develop our argument through detailed examination of the Moral Imagination methodology, a workshop-based intervention we have developed, implemented, and iteratively refined through over 70 engagements with engineering teams at Google between 2019 and 2025. Moral Imagination draws on pragmatist conceptions of ethical deliberation as imaginative inquiry \citep{Dewey1922, Werhane1999} to help teams recognize the limitations of their existing perspectives and envision alternatives. Through structured deliberation, technomoral scenarios, and facilitated perspective-taking, this methodology aims to shift team norms from exclusive focus on technical excellence toward expanded ethical awareness and responsibility. 

The paper makes three contributions. First, we introduce epistemic trust as an organizing concept for why bottom-up ethics interventions succeed or fail in engineering contexts, specified through five dimensions---Relevance, Inclusivity, Agency, Authority, and Alignment---and a model linking intervention design to engagement. Second, from systematic analysis of our facilitation experience, we build a typology of 24 failure modes organized by dimension. Third, we derive nine design principles that mitigate them. We acknowledge limitations such as selection effects from voluntary participation and the absence of formal outcome measures, and position our failure mode typology as practitioner hypotheses warranting further empirical validation.

We proceed as follows. We first situate our work within the literature on collaborative socio-technical integration and develop epistemic trust as a mechanism concept. We then present moral imagination as an intervention construct and describe the Moral Imagination program in operational detail. We next present our findings on failure modes that undermine epistemic trust and derive design principles that mitigate them. We close by discussing positioning, contribution, and limitations before concluding.

\section{Situating Moral Imagination in Collaborative Integration}

\subsection{The Integrative Collaboration Field}

Efforts to integrate ethical and social considerations into scientific and technical practices have proliferated across research and innovation policy over the past two decades \citep{Macnaghten2005, Fisher2013b, Rodriguez2013}. These are based on the premise that technical expertise, while indispensable, is inherently limited in its capacity to account for the broader societal impacts of research and innovation. Various collaborative approaches have consequently emerged to address these limitations, each embodying distinct assumptions about the nature of the problem and the appropriate mode of intervention.\footnote{For further empirical discussions of the role of teams and interpersonal dynamics in software engineering teams, see \citet{Scott2001}, \citet{Caldwell2003}, \citet{Karn2008}, \citet{Glynn2010}, \citet{Somech2013}, \citet{Robbins2015}, \citet{Gerrard2018}, and \citet{Hoffmann2022}.}

\citet{Fisher2015} provide a useful systematic mapping of this ``integrative field''. They distinguish collaborative socio-technical integration from related endeavors such as public engagement and technology assessment by three characteristics: (i) it addresses variously conceptualized socio-technical divides pertaining to expertise; (ii) it operates in close proximity to expert practices; and (iii) it seeks to meaningfully transform those practices. Within this broad category, they identify further variation along two primary dimensions: whether approaches aim to (i) enhance existing goals and capacities of focal expert practices (``native values'' and ``native capacities'') or (ii) introduce new goals and capacities for expert consideration (``alternative values'' and ``alternative capacities''). These dimensions yield four idealized modes of integration: \emph{reform} (introducing alternative values through alternative capacities), \emph{augment} (supplementing native values with alternative capacities), \emph{facilitate} (enhancing native values through native capacities), and \emph{problematize} (introducing alternative values through native capacities).

Several existing methodological frameworks instantiate these modes.

Socio-Technical Integration Research (STIR) embeds a humanist or social scientist within research laboratories to engage practitioners in structured dialogue about their decision-making. This aims to heighten reflexive awareness and broaden the considerations practitioners bring to bear on their work \citep{Fisher2006, Schuurbiers2011, Fisher2013, Smolka2021}. The Toolbox Project convenes cross-disciplinary teams for facilitated dialogue aimed at surfacing and negotiating epistemic and axiological assumptions that shape collaborative research \citep{Eigenbrode2007, ORourke2013, Looney2013, Hubbs2020}. In Value Sensitive Design, conceptual, empirical, and technical methods are applied to bring human values into technology design processes \citep{Friedman2008, Friedman2019}. Lastly, Human Practices pursues collaborative inquiry among anthropologists, biologists, and ethicists during innovation processes with the aim of working toward mutual flourishing \citep{Rabinow2009}.

These frameworks have demonstrated measurable outcomes. STIR studies have documented ``modulations''---changes in practitioners' considerations and actions attributable to engagement with embedded researchers---across diverse laboratory contexts \citep{Schuurbiers2011, Flipse2014}. Toolbox workshops have been reported by participants to improve cross-disciplinary communication and mutual understanding of collaborators' underlying research assumptions \citep{ORourke2013}. \citet{Flipse2018} demonstrate that even compressed timeframes can yield observable modulations when supported by appropriate frameworks.

Yet while the literature documents \emph{that} such interventions can succeed, it offers less systematic analysis of \emph{why} they succeed, that is the conditions under which practitioners come to regard integration efforts as credible.

\subsection{Epistemic Trust as a Mechanism}

We propose that the concept of \emph{epistemic trust} helps to fill this gap. We conceptualize epistemic trust as the degree to which practitioners regard an intervention, its facilitators, and its substantive content as worthy of serious engagement.\footnote{More generally, ``epistemic trust'' refers to an individual's willingness to consider another person's testimony as reliable and relevant. See \citet{McCraw2015}, \citet{SchroderPfeifer2018}, and \citet{Fonagy2014} for discussion.} Our conceptualization draws on philosophical analyses of testimony and expertise, which examine the conditions under which individuals rationally rely on others' knowledge in domains where they lack direct evidence \citep{Hardwig1985, Goldman1999, Origgi2004, Zagzebski2012, McCraw2015, Irzik2019} and on psychological research demonstrating that trust functions as a precondition for openness to new information \citep{Fonagy2014}.

Recent work has operationalized epistemic trust as a multidimensional construct. \citet{Hendriks2015} identify perceived competence, benevolence, and integrity as key dimensions of lay-people's trust in experts. In organizational contexts, research on psychological safety demonstrates that engaging in interpersonally risky behaviors, such as voicing dissent or admitting uncertainty, depends on trust that such behaviors will not be punished \citep{Edmondson1999}. Applied to integration interventions, these findings suggest that practitioners' engagement depends, at least in part, on whether the conditions of delivery warrant trust.

We propose epistemic trust as a \emph{mechanism} that links  intervention design to outcomes. Figure 1 below presents this pathway schematically:

\begin{figure}[htbp]
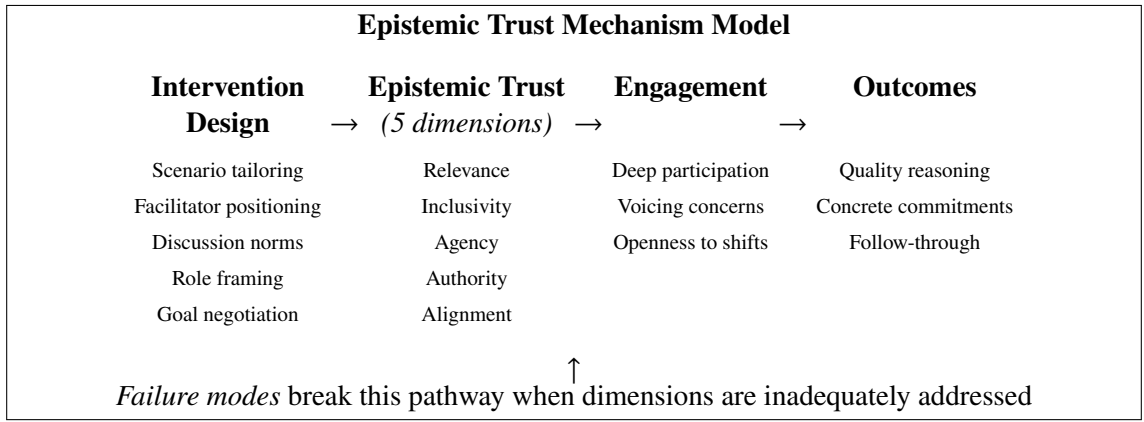

\centering
\small
\fbox{\parbox{0.9\textwidth}{%
\centering
\textbf{Epistemic Trust Mechanism Model}\\[1em]
\begin{tabular}{@{}c@{\hspace{0.3em}}c@{\hspace{0.3em}}c@{\hspace{0.3em}}c@{\hspace{0.3em}}c@{\hspace{0.3em}}c@{\hspace{0.3em}}c@{}}
\textbf{Intervention} & & \textbf{Epistemic Trust} & & \textbf{Engagement} & & \textbf{Outcomes} \\
\textbf{Design} & $\rightarrow$ & \emph{(5 dimensions)} & $\rightarrow$ & & $\rightarrow$ & \\[0.3em]
\scriptsize Scenario tailoring & & \scriptsize Relevance & & \scriptsize Deep participation & & \scriptsize Quality reasoning \\
\scriptsize Facilitator positioning & & \scriptsize Inclusivity & & \scriptsize Voicing concerns & & \scriptsize Concrete commitments \\
\scriptsize Discussion norms & & \scriptsize Agency & & \scriptsize Openness to shifts & & \scriptsize Follow-through \\
\scriptsize Role framing & & \scriptsize Authority & & & & \\
\scriptsize Goal negotiation & & \scriptsize Alignment & & & & \\
\end{tabular}\\[0.8em]
$\uparrow$\\[-0.3em]
\emph{Failure modes} break this pathway when dimensions are inadequately addressed
}}
\caption{Epistemic Trust Mechanism Model}
\label{fig:mechanism}
\end{figure}

On this model, intervention design choices shape the degree of epistemic trust practitioners extend to the intervention. For example, this may include how scenarios are tailored, how facilitators position themselves, what norms govern discussion, how participant roles are framed, and how goals are negotiated. Epistemic trust in turn enables engagement through willingness to participate, voice concerns, and remain open to perspective shifts. Engagement, in turn, produces desirable ethical outcomes such as the quality of ethical reasoning during deliberation and follow-through behavior after the intervention concludes.

We here define ``failure modes'' as design-context mismatches that break this pathway. When an intervention fails on \emph{Relevance}, practitioners disengage because they perceive ethical content as disconnected from their actual work. Authority failures are different and occur when facilitators get dismissed as lacking the competence to engage with technical realities. And when an intervention fails on \emph{Inclusivity}, minority perspectives go unvoiced and deliberation collapses into superficial consensus. These failures are observable in workshop dynamics and debrief reflections, and in our experience they accompany diminished outcomes.

Building on this model and informed by our facilitation experience, we identify the following five dimensions of epistemic trust particularly salient for collaborative ethics integration interventions in engineering contexts:

\begin{enumerate}
\item \textbf{Relevance:} This dimension addresses the challenge that ethical frameworks are often perceived as abstract or detached from technical priorities. \citet{Fisher2006} found that societal considerations are most likely to influence research practice when they are integrated at the ``midstream'' point where technical decisions are actually made as opposed to imposed externally. \citet{Fisher2013b} characterize such approaches as ``soft interventions'' that make responsible innovation operative within actual R\&D settings, though they note a persistent tension between maintaining critical distance and achieving institutional effectiveness. \citet{Balmer2016} note that Ethical Legal and Social Implications (ELSI)-style ethics work frequently fails because it positions ethics as external commentary and not as an internal contribution. Relevance is achieved in cases where practitioners perceive ethical reflection as enhancing their core objectives.

\emph{Observable indicators:} Practitioners connect ethical concepts to their own project decisions; they reference workshop content in subsequent technical discussions; and they identify ethical considerations as ``useful'' and not just ``required.''

\item \textbf{Inclusivity:} This dimension responds to the risk that deliberation becomes dominated by senior voices or collapses into superficial consensus. Research on team dynamics demonstrates that diversity of perspective improves decision quality only when team norms support the expression of minority viewpoints \citep{Caldwell2003}. Moreover, \citeauthor{Edmondson1999}'s (\citeyear{Edmondson1999}) work on psychological safety shows that raising concerns depends on confidence that doing so will not damage one's standing.

\emph{Observable indicators:} Junior team members speak substantively; disagreement is voiced and explored; and participants report feeling safe to express uncertainty or concerns.

\item \textbf{Agency:} This dimension reflects findings that top-down interventions generate resistance when practitioners feel they are passive recipients of external mandates. For instance, \citet{Lee2019} found that even when engineers engaged with ethics prompts, they frequently deflected responsibility to managers or institutional structures, suggesting learned passivity regarding ethical dimensions of their work. This pattern reflects what \citet{Basart2013} critique as the ``heroic engineer'' model, i.e. that traditional engineering ethics locates responsibility in individual practitioners, yet organizational and socio-technical realities distribute responsibility across networks of actors, making individual agency feel futile. 

\emph{Observable indicators:} Participants frame ethical considerations as ``our responsibility'' in contrast to a concern of ``management'' ; they generate their own ethical questions ; and they take ownership of commitments.

\item \textbf{Authority:} This dimension concerns the credibility of facilitators and the accessibility of ethical content. Integration research suggests that humanists and social scientists embedded in technical contexts must develop ``interactional expertise'' defined as sufficient fluency in the technical domain to engage as peers while contributing distinctive perspectives \citep{Collins2007}. For example, \citet{Lee2019} report that structured interviews yielded richer engagement when conducted by researchers who could connect ethical questions to specific technical features of participants' work.

\emph{Observable indicators:} Participants treat facilitators as credible interlocutors; they engage with ethical concepts; they accept facilitator challenges to their assumptions.

\item \textbf{Alignment:} This dimension addresses concerns that ethics interventions may be perceived as adversarial, as compliance rituals disconnected from practice, or as isolated from ongoing work \citep{vanOudheusden2014}. \citet{Flipse2018} emphasize that responsible innovation receives uptake when it demonstrably advances project objectives alongside ethical considerations.

\emph{Observable indicators:} Participants see ethical reflection as supporting their work; they integrate workshop outputs into project artefacts; follow-up actions are taken.
\end{enumerate}

These five dimensions can be illuminated by the philosophical literature on trust. Following \citet{Baier1986}, philosophers distinguish trust from mere reliance in that trust involves vulnerability and an expectation of goodwill, as opposed to just predictable competence. \citet{Scheman2001} and \citet{Goldenberg2021} extend this analysis to epistemic contexts, arguing that trust in scientific testimony depends on perceived alignment of interests between testifier and audience. On this view, Relevance and Authority function primarily as conditions for epistemic reliance and practitioners must perceive content as applicable and facilitators as competent. Inclusivity, Agency, and Alignment add distinctively trust-involving elements such as perceived goodwill, shared purpose, and conditions under which vulnerability is warranted. Our construct of epistemic trust thus synthesizes both reliance-conditions and trust-conditions as jointly necessary for genuine engagement.

Furthermore, these dimensions also map onto \citeauthor{Balmer2016}'s (\citeyear{Balmer2016}) ``rules of thumb'' for post-ELSI collaboration. Their call for \emph{collaborative experimentation} by developing integration practices jointly with practitioners instead of imposing them corresponds to our emphasis on Relevance and Agency: ethics work must be co-created with teams, adapted to their specific contexts, and positioned as enhancing their objectives. Their emphasis on \emph{taking risks} connects with Authority and Agency since facilitators must venture beyond comfortable scripts, and participants must be willing to voice perspectives that challenge team consensus. And \emph{collaborative reflexivity} aligns with our treatment of Alignment as involving transparent negotiation of goals as opposed to a form of performed consensus. Their rule to \emph{open up discussions of unshared goals} directly parallels our caution that Alignment failure modes arise when goal divergence remains unacknowledged. Finally, \emph{neighbourliness} informs our understanding of Inclusivity and Authority. It highlights that psychological safety can be a condition for determining whether collaboration enables genuine critical engagement.

Moreover, epistemic trust as we employ it is related to but distinct from several adjacent constructs. \emph{Psychological safety} \citep{Edmondson1999} concerns willingness to take interpersonal risks within a team. These include the risk of voicing dissent, admitting error, or asking a question without fear of punishment. Psychological safety is a precondition for our Inclusivity dimension but does not capture whether practitioners regard the \emph{content} of an intervention as credible or relevant. \emph{Trust in experts} \citep{Hendriks2015} concerns laypeople's willingness to defer to expert testimony based on perceived competence, benevolence, and integrity. Our Authority dimension draws on this work, but epistemic trust as we conceptualize it extends beyond facilitator credibility to include the perceived relevance of ethical content, the degree of practitioner agency, and alignment with ongoing work. \emph{Source credibility} in communication research similarly focuses on message sources. Epistemic trust therefore concerns the condition under which practitioners treat an intervention as one worth taking seriously across several dimensions at once.

\section{Moral Imagination as an Intervention Construct}

In this section, we illustrate in more detail how the moral imagination methodology can be understood as promoting epistemic trust. We first provide a high-level overview of the moral imagination methodology and then explain the application of the workshop in more detail, including relevant modules that form part of the core curriculum.

\subsection{Definition and Philosophical Rationale}

Our methodology draws on pragmatist conceptions of moral imagination, particularly John Dewey's (\citeyear{Dewey1922}) account of moral deliberation as imaginative inquiry. For Dewey, moral deliberation is an imaginative process in which agents construct and evaluate possible courses of action within problematic situations. He characterized this process as ``dramatic rehearsal'', which refers to the imaginative projection of competing lines of action and their probable consequences and which engages affective as well as cognitive capacities. On this view, moral judgment is a matter of responsiveness to the particular features of situations, informed by but not reducible to prior experience and general principles.

Subsequent pragmatist scholarship has developed Dewey's insights in directions relevant to our approach. \citet{Fesmire2003} reconstructs Dewey's ethics around moral imagination and argues that moral deliberation is a socially embedded process in which agents rehearse possible actions. He argues that moral deliberation shares the structure of artistic creation, relying on cultivable virtues such as sensitivity, discernment, creativity, foresight, and experimental intelligence \citep{Fesmire1999}. \citet{Alexander1993} makes a similar point. He argues that moral imagination involves reconstructing the meaning of a situation. \citet{Johnson1993} extends this tradition by drawing on cognitive science to challenge what he calls the ``moral law'' picture of ethics. His account emphasizes that moral imagination concerns the \emph{generation} of new possibilities, which involves envisioning actions, framings, and solutions not available before. This generative dimension is central to our methodology.

\citet{Brown2020} applies moral imagination specifically to scientific and technical practice, proposing an ``ideal of moral imagination'' that treats both inquiry and value judgment as problem-solving practices guided by imaginative evaluation of options. His framework offers resources for ethically assessing decisions in contexts where values pervasively shape inquiry. Accordingly, Brown's emphasis on moral imagination as a practice-level capacity for managing value-laden decisions in technical work directly informs our operationalization.

This pragmatist conception departs from rationalist approaches that treat moral reasoning as the application of abstract principles to concrete cases. Whereas such approaches risk generating ethical analyses disconnected from actual practice \citep{Nordmann2009}, the pragmatist tradition insists that moral reflection must be embedded in the texture of practical situations. In this context, imagination serves as a link because it allows agents to overcome their immediate perspective while remaining grounded in concrete decision-making contexts.

In business ethics, \citet{Werhane1999, Werhane2008} develops moral imagination as a practical capacity for organizational decision-making under conditions of complexity, role constraint, and bounded attention. Her central claim is that ethical failures often arise not from simple rule-breaking but from overly narrow mental models that tacitly structure what the ``problem'' is, which stakeholders are visible, what responsibilities are salient, and which options appear feasible. Moral imagination, on this account, is the disciplined ability to expose and investigate those assumptions, to re-describe situations from alternative standpoints (including the standpoints of affected stakeholders), and to generate options that are both ethically responsive and organizationally actionable.

Based on these traditions, we identify three core capacities that moral imagination involves:

\begin{enumerate}
\item \textbf{Recognition of perspectival limitation}: The ability to register that one's current framing of a situation---including what counts as the problem, who counts as a stakeholder, and what options appear available---is partial and potentially distorting.
\item \textbf{Generative envisioning}: The ability to imagine alternative framings, options, and courses of action not given in one's initial understanding.
\item \textbf{Evaluative rehearsal}: The ability to project and assess the consequences of imagined alternatives for oneself and affected others, engaging both cognitive and affective capacities.
\end{enumerate}

For the purposes of this paper, we foreground the first two capacities, which are most directly targeted by our intervention:

\begin{quote}
\textbf{Moral Imagination:} The ability to (i) register that one's perspective on a decision-making situation, including the available options and the normative factors relevant to adjudicating those options, is limited; and to (ii) creatively imagine alternative perspectives that reveal new approaches to that situation or new considerations that bear on the competing approaches.\footnote{\citet{Lange2025} provide the full conceptual account of the moral imagination methodology; see also \citet{Keeling2024}.}
\end{quote}

The first component of this account corresponds to what Dewey called the ``problematic situation'' which refers to the disruption of habitual response that initiates inquiry. Because ethical considerations rarely disturb routine practice in engineering contexts, interventions must actively create conditions for such disruption. The second component requires structured support, since imagination detached from disciplined inquiry risks both confabulation and conservatism.

\subsection{Positioning Within Integration Approaches}

Our methodology shares similarities with, but also differs from, other integration approaches. Like STIR's use of a decision protocol to surface considerations informing research choices \citep{Fisher2013}, Moral Imagination employs structured questioning to make implicit reasoning explicit. Moreover, similar to the Toolbox Project's facilitated dialogue \citep{ORourke2013, Crowley2010}, it creates dedicated space for articulating assumptions that normally remain tacit.

Following \citeauthor{Fisher2015}'s (\citeyear{Fisher2015}) taxonomy, Moral Imagination can be characterized and distinguished from existing approaches more precisely as follows:

\emph{Forms:} The approach is embedded \textit{within} organizational practice. It represents societal dimensions as latent in engineering work, which positions it primarily in the ``problematize'' mode (introducing alternative values through native capacities) while drawing on ``augment'' elements when introducing ethical concepts unfamiliar to participants.

\emph{Means:} The approach is semi-structured. Technomoral scenarios are tailored to specific project contexts, but the deliberation protocol (Reflect, Articulate, Action) provides consistent scaffolding. Facilitators may decide to spend more or less time on particular topics, depending on the need for specific ethical topics to be explored for a given participating team. This resembles STIR's balance between structured decision protocol and responsive engagement.

\emph{Ends:} The approach seeks normative transformation, specifically, expanded perception of role obligations among engineering practitioners. This positions it toward the ``alternative values'' pole, though it pursues this transformation by engaging capacities (imagination, deliberation) that are native to engineering practice.

However, while STIR typically involves extended embedding of a single researcher in laboratory practice over weeks or months, and Toolbox convenes one-time workshops focused primarily on epistemic assumptions, Moral Imagination combines intensive workshop engagement with explicit follow-up mechanisms and forward-looking action planning.\footnote{A dual approach like this has precedent in responsible innovation. \citet{Smolka2023}, for instance, adapt STIR for innovation-ecosystem governance by curating recurring workshops, training participants as ``brokers'' who carry the protocol into their local work contexts, and pairing these with stakeholder and scenario workshops.} This positioning reflects a strategic judgment about the conditions of epistemic trust in corporate engineering contexts, in which extended embedding may be impractical and where connection to actionable outcomes is essential for perceived relevance.

\section{The Moral Imagination Program}

\subsection{Overall Objective}

The main objective of the Moral Imagination workshop is to shift a team's understanding of their responsibilities from (a) their being responsible for the merely technical aspects of their work to (b) their being responsible for the ethical and sociotechnical aspects of their work. This shift in understanding is realized through a change in team norms; that is, from the standard norms of engineering culture to a more expansive set of norms that emphasize ethical awareness, deliberation, and decision-making. Thus a central aim of our approach is to facilitate a \emph{role obligation shift} among engineers. To facilitate this shift, the workshop helps the team to develop and promote a culture of ethical awareness, deliberation, and action in which values relevant to the team's work and possible harms arising from the work are articulated, discussed, and well-understood. Lastly, the workshop aids teams in developing ethical commitments and a roadmap that captures the role of ethical justification in technical design decisions going forward \citep[cf.][on strengthening organizational ethical capabilities]{Lange2025}.

Crucial to this capability is ``becoming aware of one's context, understanding the conceptual scheme or `script' dominating that context, and envisioning possible moral conflicts or dilemmas that might arise in that context or as outcomes of the dominating scheme'' \citep[p.~3]{Werhane2008}. Moral Imagination consequently allows engineers to recognize the limitations of their pre-theoretic mental models about how their technology impacts the world, what the costs and benefits of that technology are, and what \emph{their} role is in ensuring responsible technological development \citep{Haidt2001}. It aims to shift teams' self-conception away from a mindset where ethical considerations are removed from perceived responsibilities---something that ``falls outside of the job description''---toward a mindset where the consideration of the moral implications is an inherent part of the research and development process.

In practice, this entails explicitly inviting teams to question and potentially reframe their project goals, problem definitions, and success metrics. The process encourages participants to interrogate the fundamental premise of their work, asking whether they are solving the right problem, for whom, and under what assumptions. This critical inquiry empowers teams to move beyond simply mitigating risks within an existing plan, make hard decisions together, and to determine whether certain aims should be deprioritized or, in some cases, not pursued at all.

\subsection{Facilitation Objectives}

Overall, our methodology pursues three types of facilitation objectives that run in tandem throughout the workshops, reflecting distinct but complementary design aims.

First, \emph{ethical objectives} introduce substantive concepts from moral philosophy and responsible innovation research such as stakeholder analysis, value trade-offs, and anticipatory reasoning, which scaffold teams' ethical understanding \citep{Owen2013, Stilgoe2013}. Second, \emph{applied objectives} focus on connecting these concepts to the team's specific work context, addressing the well-documented challenge of operationalizing abstract ethical principles in concrete technical practice \citep{Fisher2006, vandePoel2013, Lee2019}. Third, \emph{experiential objectives} ensure that participants can meaningfully engage with both the ethical and applied content through structured activities that foster psychological safety, perspective-taking, and constructive disagreement \citep{Edmondson1999, Caldwell2003}.

This tripartite framework has proven useful for iterating workshop content: it clarifies which aspects of the methodology are working well and which require refinement. Together, these objectives aim to support the three ethical capabilities introduced above: building ethical awareness, ethical deliberation and decision-making, and ethical commitment (Table~\ref{tab:objectives}).

\clearpage
\begin{table}[h!]
\centering
\caption{Moral Imagination Facilitation Objectives}
\label{tab:objectives}
\footnotesize
\renewcommand{\arraystretch}{1.4}
\begin{tabular}{@{}p{2.8cm}p{3.8cm}p{3.8cm}p{3.8cm}@{}}
\toprule
& \textbf{Ethics Objectives} & \textbf{Applied Objectives} & \textbf{Experiential Objectives} \\
\midrule
\textbf{Building Ethical Awareness} & Awareness Building for Values in Technology: Sensitize individuals and teams to the importance of ethical reflection and enhancing their understanding of the potential impact of values on technology design decisions. & Understanding Implicit Values in Group Dynamics: Align the team's values, perspectives, and aspirations through tailored learning methods while identifying potential value conflicts in their work and/or product. & Fostering Engagement and Trust: Encourage active participation, build psychological safety, and practice constructive disagreement and exploring diverse perspectives. \\
\midrule
\textbf{Ethical Deliberation, Reasoning, Decision-Making} & Ethical Deliberation and Decision-making: Help teams with ethical deliberation, including understanding various moral considerations, stakeholder perspectives, potential conflicts, and societal implications.& Ethical Considerations in the Day-to-Day: Help teams understand how ethical principles and considerations apply to their work and how teams can make informed and responsible decisions, navigate complex moral dilemmas, and consider the perspectives of various stakeholders in these decision scenarios. & Engagement Through Empathy and Diversity of Perspectives: Foster empathy for diverse perspectives, encourage open dialogue about potential risks, and cultivate a nuanced understanding of ethical dilemmas in the real world.\\
\midrule
\textbf{Ethical Commitment, Action, and Follow-Through} & Moral Courage: Help teams act as the kind of person who is naturally inclined to pursue a morally informed course of action. & Embracing Responsibility: Through reflective discussions and a focus on the team's specific technology or product, participants identify the most salient ethical issues, prioritize areas for resource allocation, and develop a tangible artefact to guide future decision-making. & Engagement through Empathy and Diversity of Perspectives: Foster empathy for diverse perspectives, encourage open dialogue about potential risks, and cultivate a nuanced understanding of ethical dilemmas in the real world. \\
\bottomrule
\end{tabular}
\end{table}

\subsection{Design Logic and Workshop Structure}

The Moral Imagination program operationalizes the concept developed above through a structured modular workshop format designed to cultivate both components of moral imagination, namely recognition of perspectival limitation and creative envisioning of alternatives. Workshops are typically conducted over one to two days, depending on team availability and depth of engagement desired.

To begin, participation is voluntary. Teams self-select into the program; they are not required to participate as a compliance or review mechanism. Workshops are typically initiated by a team lead, product manager, or team member who perceives value in structured ethical reflection on their work. This voluntary uptake model reflects a deliberate design choice. Mandatory ethics interventions risk triggering the ``external imposition'' failure mode discussed below; voluntary participation establishes baseline alignment and signals that the workshop serves the team's own purposes. However, voluntary participation also introduces selection effects since teams that engage are likely more receptive to ethical reflection than the engineering population at large, and our findings should be understood in this light.

Workshops are organized around three phases: (i) Reflect, (ii) Articulate, and (iii) Action \citep[cf.][]{Owen2013}.

\textbf{Reflect Phase:} The workshop opens with exercises designed to uncover participants' existing assumptions about their technology's impacts and their responsibilities as developers. Facilitators introduce a technomoral scenario (TMS). This is a narrative projecting the team's technology into a future context where its ethical implications become salient. Scenarios are tailored to each team's specific project during intake scoping discussions, drawing on potential use cases, user populations, deployment contexts, and foreseeable misuses. The scenario functions as a Deweyan ``problematic situation'' since it presents a disruption of routine perspective that initiates inquiry.

Participants engage with the TMS through structured role-play. They are assigned to represent different stakeholders such as users, affected communities, regulators, adversarial actors, and then asked to articulate how the technology appears from each perspective. This perspective-taking exercise is designed to trigger recognition of perspectival limitation and recognition that participants' default engineering perspective occludes considerations visible from other positions \citep[see][for an overview of TMS construction and implementation]{Keeling2024}.

\textbf{Articulate Phase:} Having surfaced the limitations of their existing perspective, participants move to structured deliberation about the ethical considerations raised by the scenario. Facilitators guide discussion through a protocol that asks: What values are at stake? Whose interests are affected? What responsibilities do we have? What are we uncertain about?

The Articulate phase aims to move beyond individual perspective-taking to collective deliberation. Participants are asked to reason together about the relative importance of respective ethical considerations, potential conflicts, and implications for design decisions. Facilitators introduce ethical concepts (e.g., fairness, autonomy, dignity, justice) as tools for articulating intuitions that participants may struggle to express in their native technical vocabulary.

\textbf{Action Phase:} The workshop concludes with forward-looking planning. Participants are asked to translate their deliberations into concrete commitments: Responsibility Objectives that can be incorporated into project planning artefacts such as OKRs (Objectives and Key Results) or PRDs (Product Requirements Documents). This phase addresses a common failure mode of ethics interventions: generating insight without enabling action.

Responsibility Objectives are intended to be specific and integrate with existing project-management structures. They specify particular investigations, design modifications, or review processes that the team commits to undertake. This integration with project artefacts is essential for perceived relevance and follow-through.

\subsection{Intake, Facilitation and Follow-Up}

Each workshop is preceded by a scoping discussion with the team member or lead who initiated participation. This discussion serves multiple functions: (i) gathering information needed to tailor the technomoral scenario; (ii) identifying team dynamics that may affect deliberation; (iii) establishing facilitator credibility; (iv) negotiating workshop objectives; and (v) confirming that participation reflects real team interest. The intake process is important for Relevance and Authority: without understanding a team's specific project context and motivations for engaging, facilitators cannot craft scenarios that connect to participants' actual work and concerns.

Facilitators are trained to balance structure with responsiveness. They provide scaffolding for deliberation while remaining open to directions that participants find meaningful. Facilitators are selected for dual domain expertise. They need to have familiarity with ethical concepts and frameworks, combined with sufficient technical background to engage credibly with engineering teams. This dual expertise aims to address the Authority dimension since facilitators must be able to translate between ethical and technical domains, by, for example, connecting abstract concepts to specific design decisions. Moreover, facilitators are trained to cultivate psychological safety, particularly through modeling intellectual humility and treating participant contributions with  curiosity. They are also trained to ``challenge benevolently'' by pushing participants beyond their comfort zones while maintaining trust that the challenge is offered in the spirit of shared inquiry.

Workshops are followed by debrief discussions with the facilitation team, assessing what worked and what didn't. Where possible, facilitators conduct follow-up with teams to assess whether Responsibility Objectives were enacted. This follow-up data, while not systematically collected, informs iterative refinement of the methodology.

\section{Findings: Failure Modes That Undermine Epistemic Trust}

This section presents our central empirical contribution: a typology of failure modes that undermine epistemic trust in bottom-up ethics interventions. These findings emerged from systematic analysis of facilitator debriefs, workshop notes, and structured interviews conducted across more than four years and over 70 workshops. We organize the findings by the five dimensions of epistemic trust introduced above, then identify cross-cutting failure modes that involve interactions between dimensions.

A preliminary note on scope. As noted in the previous section, participation in Moral Imagination workshops is based on self-selection. Our findings therefore describe failure modes that arise even among teams with prior openness to ethics engagement. These challenges persist despite favorable initial conditions; they are not artifacts of forcing resistant teams to participate. Conversely, we have no data on teams that decline to engage, and our findings should not be generalized to contexts where participation is mandatory.

\subsection{Methodological Approach}

Our methodology is best characterized as a practice-based qualitative inquiry aimed at developing analytically generalizable insights from an extended program of ethics facilitation in organizational settings. Our contribution is therefore not a formal outcome evaluation of the intervention, but a systematic analysis of the conditions under which ethics workshops are taken up, resisted, dismissed, or co-opted by engineering practitioners. We report a typology of failure modes and associated design principles as practitioner hypotheses grounded in qualitative analysis of intervention artefacts and accounts of workshop dynamics.

Between 2019 and 2025, the facilitation team delivered 70+ Moral Imagination workshops to product and research teams, engaging approximately 1000 participants. Participation was voluntary, and teams self-selected into the program. Each workshop was preceded by an intake scoping discussion with initiating team members to tailor the technomoral scenario and clarify the team's motivations and constraints, and followed by structured facilitator debrief discussions.

The analyzed data consisted of three primary categories of material: (i) intervention artefacts (e.g., agendas and slide decks, technomoral scenario, interactive workshop worksheets and/or digital boards where used, responsibility-objective drafts, and documented action plans); (ii) contemporaneous records of workshop dynamics (facilitator notes and pre- and debrief memos capturing points of friction, participant responses, discussion trajectories, and follow-up actions); and (iii) retrospective elicitation materials collected by the first author, including non-participant observation field notes from 10 workshops, semi-structured interviews with primary facilitators, semi-structured interviews with a subset of workshop participants, and group interviews with the facilitation team aimed at drawing out recurring patterns and boundary conditions. Where available, follow-up communications (e.g., stakeholder check-ins and references to workshop outputs in subsequent project discussions) were used to corroborate the translation of workshop insights into ongoing work.

Given the scale and heterogeneity of the program, we did not treat all workshops as equivalent units for in-depth coding. Instead, we combined (a) maximum-variation purposive sampling for deep analysis---i.e. prioritizing workshops with rich documentation and variation across team type, seniority mix, project maturity, and organizational context---with (b) confirmatory scanning across the broader workshop set to test whether emergent patterns recurred beyond the deep-coded subset and to identify exceptions. This approach was selected to balance breadth (capturing variation across contexts) with analytic depth (preserving  detail needed to identify mechanisms of engagement and failure).

Our analysis proceeded abductively in two stages. Stage 1 (mapping and incident extraction) involved producing structured analytic memos for each workshop in the deep-analysis subset and extracting ``critical incidents'': discrete episodes in which (i) ethical reflection was substantively taken up, (ii) discussion stalled or was resisted, (iii) the intervention was framed as irrelevant/adversarial/inactionable, or (iv) responsibility was displaced or domesticated into superficial commitments. These incidents served as the primary unit of analysis because failure modes are expressed most clearly in the micro-dynamics of facilitation, participation, and contestation. Stage 2 (incident-based coding and synthesis) involved coding incidents using a small codebook that was organized around the five epistemic-trust dimensions as sensitizing concepts, alongside inductive codes capturing recurring interactional mechanisms (e.g., boundary policing and role deflection, goal foreclosure, demands for legibility/metrics, authority deferral, time-pressure vetoes). Codes were iteratively refined through analytic memoing, comparison across incidents, and clustering into candidate failure modes. The failure-mode typology represents the stabilized outcome of this iterative process, and the design principles represent design implications drawn from recurrent relationships between failure modes and facilitation choices.

We strengthened credibility through triangulation and iterative challenge. First, patterns were triangulated across data types (artefacts, contemporaneous notes/debriefs, observations, participant interviews, and facilitator interviews) to reduce reliance on any single source or perspective. Second, we actively searched for incidents that did not fit emerging failure-mode definitions and revised definitions accordingly (e.g., identifying the conditions under which role-play was accepted, or when alignment enabled rather than neutralized critical challenge). Third, failure-mode definitions and principle mappings were subjected to peer debriefing in structured group discussions with the facilitation team, which served both to pressure-test interpretive claims and to identify alternative explanations rooted in context (e.g., team lifecycle, deadline pressure, leadership dynamics). Finally, we note the interpretive risks inherent in this practice-based inquiry. Since facilitators are stakeholders in the intervention's success, we treat the typology as analytically generalized as opposed to universally validated, and we explicitly delimit the scope introduced by voluntary participation and the absence of formal outcome measures (see the Limitations below).

\subsection{Failure Modes by Epistemic Trust Dimension}

\subsubsection{Relevance Failures}

When interventions fail on \emph{Relevance}, practitioners disengage because they perceive ethical content as disconnected from their work.

\emph{External imposition or punitive framing.} Participants perceive the workshop as something done to them, or feel their work is under attack. This failure mode arises even with voluntary participation when not all team members were involved in the decision to participate, or when organizational context shifts between sign-up and delivery. Facilitator debriefs identified this dynamic particularly in larger teams (e.g. 20+ participants), where participation quality could be lower and commitments remained more superficial compared to recommended group sizes of 8 to 16 participants. Relatedly, when facilitators are perceived as having a pre-determined agenda (``Are you imposing your values on us?'') or judgmental attitude (``Are you going to block us for saying this?''), teams become defensive.

\emph{Abstraction without connection.} Ethical concepts are introduced at a level of generality that participants cannot connect to concrete design decisions. Discussions of ``fairness'' or ``autonomy'' remain at the level of principle without translation to specific technical choices. This might reflect prior corporate ethics training participants had received, or the course on philosophy one participant had in college. In interviews, participants with that background often doubted that any philosophical approach was relevant for their work. The pattern was concentrated in early workshops, before we had protocols for grounding abstract concepts in project-specific examples.

\emph{Scenario irrelevance.} The technomoral scenario fails to connect to the team's specific project context. This could occur through inadequate intake scoping, reliance on generic scenarios, or projecting so far into the future that participants cannot recognize their technology. Comparison across incidents indicated that a trade-off exists between how far into the future the scenario authors ideate, including the technology's impacts on social dynamics, and the relevance the scenario holds to participants' day-to-day decision-making. Across the interviews, scenario tailoring stood out as the single highest-leverage factor in a workshop's success.

\emph{Conversational drift.} Without consistent guardrails set by facilitators, discussion about ethical values, which are vital for the workshop, may become focused on participants' personal values and experiences and not on the implication of technical product decisions over time. When this occurs, the intervention loses its connection to technical decision-making and participants question the relevance of their efforts in the workshop to their professional responsibilities. In several observed sessions, opening exercises invited too much personal reflection and never anchored it to the technology at hand. Debriefs confirmed this as a recurring risk.

\subsubsection{Inclusivity Failures}

\emph{Inclusivity} failures concern cases where deliberation becomes dominated by senior voices or collapses into superficial consensus.

\emph{Hierarchy reproduction.} Existing team hierarchies are reproduced in workshop dynamics, for example, when senior engineers or managers dominate discussion while junior members defer. Based on the facilitators' observations, the pattern was especially pronounced in teams with strong technical leads. Voluntary participation does not flatten hierarchy; when a senior member initiates participation, their implicit endorsement may amplify their discursive dominance. Relatedly, when technical expertise is treated as the primary qualification for ethical input, insights from non-technical team members (designers, product managers, user researchers) are devalued or ignored. In the group interviews, the facilitation team returned to this as a recurring concern, one requiring active mitigation.

\emph{Premature consensus.} The team converges quickly on a shared view without exploring disagreement. This may reflect conflict aversion, time pressure, monocultures of engineering teams, or facilitator failure to probe for dissent. The debrief notes mark this as recurring in teams with strong collaborative or hierarchical cultures, where disagreement carried a social or professional cost. Paradoxically, teams that worked well together sometimes produced thinner ethical deliberation. However, follow-up interviews with participants suggested that teams where participants were most comfortable expressing their views developed the most elaborate ethical frameworks and communicated those actively to their peers in other, related teams.

\emph{Role-play resistance.} Participants become uncomfortable or dismissive when scenario-based exercises feel artificial, childish, or implausible. In the facilitator interviews, the resistance came mostly from senior engineers who found role-play a silly exercise, and from juniors uncomfortable taking on a different role in a professional setting. When not properly framed, the methodology's role-play technique becomes a barrier to engagement. However, analysis of the debriefs and follow-up interviews with participants suggests that, in most cases, it only took one participant to embrace their temporary role, which would help others to play along.

\subsubsection{Agency Failures}

In \emph{Agency}-specific intervention failures, participants feel lectured to and not considered as active deliberators.

\emph{Passive reception.} Participants adopt a receptive stance and listen rather than generate their own ethical questions. The debrief discussions record this more often when facilitators over-structured the discussion, or framed too heavily before inviting input. Even self-selecting teams slip into passive mode if workshop structure positions them as the audience. In coding, the main triggers were dwelling on the history of philosophical ideas, leaning on the vocabulary of ethical theory, and expecting constructive argument from the start.

\emph{Responsibility deflection.} Participants engage with ethical considerations but deflect responsibility to other actors: managers, legal teams, policy-makers, users. This pattern, documented by \citet{Lee2019}, appeared sometimes in coded incidents. On this point, facilitator notes recorded characteristic phrases (``that's a policy decision,'' ``legal would need to weigh in,'' ``but what if users choose to accept the terms''). Voluntary participation indicates willingness to engage, though it does not automatically undo professional socialization that locates ethical responsibility elsewhere.

\emph{Calibration failures.} When ethical dilemmas are presented in oversimplified terms, sophisticated participants disengage because the framing fails to honor the complexity of day-to-day decision making realities. Non-participant observation captured episodes where facilitators were met with blank stares following oversimplified framings. Conversely, when scenarios are too complex or distant from current work, participants struggle to engage productively, however hard they may try. Facilitator interviews identified calibrating complexity to audience expertise as a key skill, with both over- and under-specification producing disengagement.

\emph{Facilitation rigidity or drift.} Two opposite failure modes emerged around facilitation structure. When facilitators hold rigidly to predetermined structure despite participant signals that it feels irrelevant, teams disengage from what feels like a dogmatic exercise. When facilitators allow too much flexibility, workshops become unfocused and unproductive, with participants uncertain what they are meant to accomplish. Group interviews with the facilitation team emphasized that judging when sufficient philosophical depth was achieved before moving on to subsequent exercises is a learned pedagogical skill. Debrief discussions corroborated that navigating between these poles required real-time judgment about when to adapt and when to maintain direction.

\subsubsection{Authority Failures}

When interventions fail on \emph{Authority}, participants dismiss facilitators or ethical content as lacking credibility.

\emph{Technical incredibility.} Some facilitators lacked sufficient technical background to engage credibly with engineering teams. In several critical incidents a facilitator's technical misstatement visibly shifted the room---for example, when a superficial reading of sophisticated machine-learning concepts was offered as a solution, though literature and practice had moved on from those ideas years earlier. Recovery required explicit acknowledgment of the limits of understanding, and often relied on co-facilitators with stronger technical backgrounds to clarify the ethical relevance of the idea suggested by their peer. Self-selecting teams expect more, participant interviews suggest, precisely because they have chosen to invest their time.

\emph{Idiom dominance.} This failure involves one idiom crowding out the other. When technical jargon dominates, ethical analysis moves to the background as participants compare solutions to technical analogies and precedents. When ethical or philosophical jargon dominates (e.g. using formal definitions of ethical theories, including words like ``deontology'' or ``virtue''), technical team members feel alienated by abstract vocabulary. Early workshops, by the facilitators' account in interview, erred toward philosophical precision at the cost of accessibility. Comparing incidents, and drawing on informal consultation with a philosopher, we shifted the protocols toward plain-language presentations that carried the same meaning for non-philosophers.

\emph{Perceived na\"{i}vet\'{e}.} Facilitators are perceived as naive about organizational and commercial realities, for example, the pressures, constraints, and incentives shaping engineering decisions. This is because participants typically operate within a web of stakeholder demands, interests, and incentives, ranging from personal preferences of their superiors, commercial market realities, a shifting policy landscape, and divergent team and/or organizational goals. This failure mode arose particularly with experienced teams under significant resource or timeline pressure; debrief memos emphasized the importance of acknowledging constraints explicitly.

\emph{Challenge perceived as attack.} When facilitators push participants to examine assumptions, challenges are perceived as personal criticism. Participants could become defensive, insecure in their understanding of the workshop content, and therefore shut down. Informed by the group interviews, the training protocols evolved to frame challenges as shared inquiry and to model intellectual humility, yet this remained among the most delicate failure modes to deal with.

\subsubsection{Alignment Failures}

These involve participants who perceive ethics opposed to their objectives or disconnected from ongoing work.

\emph{Adversarial framing.} Ethics is perceived by some participants as a constraint on innovation that impedes goals. In the facilitator interviews, such participants had often just been through a mandatory ethics training, or had had their work put through an ethics review. However, their experience had been disappointing or the recommendations impossible to apply. In coding, these associations had to be countered early, through framing that stressed shared inquiry. Once understood by one participant, a cascading effect could be noticed where others accepted their peer's reasoning and overcame their initial hesitations.

\emph{Temporal discounting.} Participants prioritize immediate deadlines and deliverables over long-term ethical implications. When teams are under delivery pressure, ethical considerations that don't bear on the current sprint feel like luxuries they cannot afford. The follow-up communications show that without someone on the team naming the cost of ethical failure against the deadline, participation tended to be postponed exactly when it mattered most. The debrief memos place this in teams approaching a major launch or facing resource constraints.

\emph{Goal divergence.} Facilitators' goals conflict with team priorities, and this divergence is not acknowledged. Teams may self-select for varied reasons, including genuine interest, curiosity, social pressure, team-building, and these motivations shape expectations. Comparison of intake documentation with debrief discussions identified unacknowledged goal divergence (e.g. between what was discussed prior to the workshop and what the participating team members hoped to address) as an occasional source of friction. When surfaced during intake meetings, it became manageable, but when latent, it undermined engagement throughout.

\emph{Isolation from practice.} Workshop insights might remain isolated from ongoing work, when facilitators do not push a team to consider what a useful outcome artefact would be. Even when deliberation produces relevant ethical insights, the nature of the artefact is key for lasting change. A failure to connect with project artefacts means insights decay. This proved a primary determinant of lasting impact, as the follow-up communications and stakeholder check-ins show. Workshops that produced concrete Responsibility Objectives integrated into Objectives and Key Results (OKRs) or Product Requirements Documents (PRDs) showed evidence of sustained attention in subsequent project discussions, while those ending in general discussion typically left little trace.

We note that Alignment raises a well-known risk in responsible innovation: interventions can be co-opted into compliance or reputation rituals. We return to this structural tension in the Discussion below.

\subsection{Cross-Cutting Failure Modes}

Several failure modes involve interactions between dimensions, making them challenging to diagnose.

\emph{Credibility-connection trap (Authority $\times$ Relevance).} Facilitators with strong technical backgrounds may achieve \emph{Authority} but struggle with \emph{Relevance} if technical fluency leads them to accept engineering framings uncritically. Conversely, facilitators who challenge assumptions may establish Relevance of ethical considerations while undermining Authority. Facilitator interviews captured this tension; one facilitator described successful navigation as ``earning the right to push back.''

\emph{Participation paradox (Inclusivity $\times$ Agency).} Structured exercises ensuring \emph{Inclusivity} can undermine \emph{Agency} if participants experience them as forced. Non-participant observation documented episodes where explicit calls on quiet participants backfired, producing reluctant minimal responses. The challenge is designing structures that enable voice without scripting it.

\emph{Co-optation risk (Alignment $\times$ Agency).} Establishing \emph{Alignment} by connecting ethics to team objectives can undermine \emph{Agency} if it positions ethical reflection as serving management priorities and not practitioners' own responsibility. When ethics becomes primarily instrumentalized as ``good for the product,'' it may lose capacity to challenge product direction. Group interviews with the facilitation team identified this as a persistent structural tension. The voluntary nature of participation may partially mitigate but not fully eliminate the failure mode.

\emph{Expertise hierarchy (Authority $\times$ Inclusivity).} Facilitators with strong expertise may inadvertently reproduce hierarchy by positioning themselves as authorities whose views carry special weight, silencing participant contributions. Memos from workshops with technically expert facilitators flagged this pattern. Training protocols evolved to emphasize facilitator questions over claims, and to explicitly invite disagreement with facilitator perspectives.

\subsection{Illustrative Episodes}

The failure modes presented above emerged from analysis of different workshop dynamics. The following vignettes illustrate how these failures manifested in practice and how facilitators attempted to address them. Each vignette has been anonymized to protect participant confidentiality.

Note that these episodes are not intended as proof of the failure mode typology but as perspectives into the workshop dynamics from which that typology emerged. They illustrate the concrete challenges that prompted our iterative refinement of the methodology and the reasoning that connects failure modes to design principles.

\begin{quote}
\textbf{Vignette 1: ``Ethics is subjective'':} In a workshop with a team working on core computing infrastructure, several participants doubted that academic ethics bore on their technical work. One engineer stated that ``ethics is subjective,'' and asked whether the facilitators were there to impose external constraints on the design process. The resistance, facilitators later gathered, stemmed from prior experiences with corporate training that felt abstract and disconnected from engineering reality. Participants perceived the workshop as top-down judgment.

Recognizing this, the facilitators paused and made the approach more explicit. They described the methodology as non-punitive and bottom-up, and recast the technomoral scenario as a space for the team to examine its own design choices. The room dynamic subsequently changed. Participants who had been guarded engaged more readily in the exercises that followed. Alignment, the episode shows, can fail before a workshop begins based on prior institutional experience. Explicit framing can mitigate it.

\emph{Relevant failure modes: External imposition or punitive framing; Adversarial framing}
\end{quote}

\begin{quote}
\textbf{Vignette 2: Technical misstatement:} Halfway through a workshop with an experienced engineering team managing a complex networked system, the lead facilitator made a technical misstatement about the system's architecture. Team members exchanged glances and several visibly withdrew. The error exposed a gap between the facilitator's understanding and the team's own familiarity with the system. 

The co-facilitator, who possessed stronger technical expertise in the domain, stepped in. Rather than glossing over the mistake, she validated the team's perspective, acknowledged the misstatement directly, and carefully explained how the ethical point her colleague had been making still applied within the corrected technical framing. The combination of epistemic humility and genuine engagement with the team's expertise was well received. Participants re-engaged, and the session surfaced ethical risks specific to the system's technical dependencies. The episode points to two things at once: how fragile Authority is, and how much a facilitation team with complementary expertise can do to protect it. In other cases, similar dynamics proved more resistant to intervention and not all Authority failures were recoverable within a workshop.

\emph{Relevant failure modes: Technical incredibility; Perceived na\"{i}vet\'{e}}
\end{quote}

\begin{quote}
\textbf{Vignette 3: Manager who spoke first:} In a workshop on a team's decision-making processes, a senior manager took control of the conversation early. He framed the project's ethical considerations as justifications for decisions already made, and directed his remarks to the facilitators. The junior members stayed largely silent and deferred. The pattern intensified when he posted reflections to the shared chat during the time designated for individual reflection, signaling that his perspective was the one that counted. 

Facilitators intervened to reshape the discussion. They restructured the next exercise to collect written responses before verbal sharing, directly invited contributions from junior team members by name, and gently redirected when the manager began speaking for others. By the afternoon, the junior engineers were voicing concerns that challenged his earlier points. A single narrative had given way to a more diverse exchange, and ethical considerations that had gone unspoken now surfaced more readily.

\emph{Relevant failure mode: Hierarchy reproduction}
\end{quote}

\begin{quote}
\textbf{Vignette 4: Abandoning the agenda:} A product team was managing a significant architectural shift in its technology stack, and the workshop's planned agenda began to feel disconnected from its immediate concerns. Participants described the exercises as a rigid checklist that did not address what they were grappling with. Engagement flagged, and the discussion grew more superficial.

The facilitators judged, in the moment, that the plan for the workshop had to change. They paused the sequence and asked the team directly which technical challenge most concerned them. The question surfaced a tension the team had been avoiding, between scaling requirements and fairness considerations. The rest of the session addressed that dilemma specifically and engagement returned once the workshop met the team's actual priorities. Participants identified a core value conflict in their scaling logic and committed to revised Responsibility Objectives for the next development cycle. 

\emph{Relevant failure modes: Facilitation rigidity; Scenario irrelevance}
\end{quote}

\begin{quote}
\textbf{Vignette 5: When the roadmap changed:} Shortly after a workshop with a team developing a business-to-business product feature, an unexpected corporate restructuring altered the project's scope. The Responsibility Objectives (ROs) the team had drafted now described a project whose scope had since changed.

The workshop's effect proved nonetheless resilient. Facilitators had encouraged the team to produce outputs beyond formal documentation: a presentation for company-wide circulation and an internal website for discussion. The original ROs were eventually shelved while these cultural artefacts were not. The presentation drew interest from adjacent teams, and the website sustained a conversation about the ethical implications of the product area. Alignment, the episode suggests, is sometimes reached through indirect channels when formal integration fails.

\emph{Relevant failure mode: Isolation from practice}
\end{quote}

\section{Design Principles: Mitigating Failure Modes}

Based on the failure mode typology, we derive design principles for cultivating epistemic trust. These consolidate best practices developed through facilitation experience. Here they are presented as design implications of the mechanism model.

\subsection{Principles Overview}

\begin{quote}
\textbf{Principle 1: Contextual Tailoring:} Ethical content must be tailored to the specific project context. It should draw on particular technical features, use cases, and deployment contexts. Scenarios should be grounded enough to feel recognizable yet projective enough to reveal non-obvious implications. Discussion must remain largely anchored to technology implications. \emph{Addresses:} Abstraction without connection; Scenario irrelevance; Conversational drift; Calibration failures
\end{quote}

\begin{quote}
\textbf{Principle 2: Non-Adversarial Framing:} Position the workshop as collaborative dialogue and not as a corrective exercise. Ethics should be presented as a resource for achieving team objectives well, not as external judgment or constraint on innovation. Attention to how participation is initiated by ensuring that team members understand and endorse the decision to engage can achieve this. \emph{Addresses:} External imposition or punitive framing; Adversarial framing
\end{quote}

\begin{quote}
\textbf{Principle 3: Dual-Expertise Facilitation:} Facilitators must operate fluently in both technical and ethical subject matter domains. This prevents either technical or philosophical vocabulary from dominating discussion and alienating participants. And co-facilitation with complementary expertise can mitigate individual limitations. \emph{Addresses:} Technical incredibility; Idiom dominance
\end{quote}

\begin{quote}
\textbf{Principle 4: Structured Voice Distribution:} Deliberation structures must actively distribute speaking opportunities, legitimate diverse contributions, and explicitly invite dissent. This counteracts default hierarchies and prevents premature consensus, so that non-technical voices are heard. \emph{Addresses:} Hierarchy reproduction; Premature consensus; Participation paradox (partial); Expertise hierarchy (partial)
\end{quote}

\begin{quote}
\textbf{Principle 5: Deliberative Ownership:} Participants must be positioned as the primary deliberators and locus of ethical responsibility for their work. Facilitators should provide scaffolding and adapt the structure to participant needs while maintaining productive direction. This counters both passive reception and the tendency to deflect responsibility elsewhere. \emph{Addresses:} Passive reception; Responsibility deflection; Facilitation rigidity; Facilitation drift; Participation paradox (partial); Co-optation risk (partial)
\end{quote}

\begin{quote}
\textbf{Principle 6: Benevolent Challenge:} Facilitators should challenge assumptions and push beyond comfort zones while, at the same time, maintaining relational trust. Challenges, if made, should be positioned as collaborative inquiry. To this end, facilitators can display intellectual humility, acknowledging the constraints participants face, and inviting reciprocal challenge. \emph{Addresses:} Challenge perceived as attack; Perceived na\"{i}vet\'{e}; Credibility-connection trap (partial); Expertise hierarchy (partial)
\end{quote}

\begin{quote}
\textbf{Principle 7: Goal Transparency:} Workshop objectives must be negotiated explicitly with teams during intake. Divergence between facilitator and team goals must be surfaced and addressed. Goals may be reformulated during an engagement, but, if so, need to be articulated. Facilitators should articulate how ethical reflection connects to team priorities. \emph{Addresses:} Goal divergence; Co-optation risk (partial)
\end{quote}

\begin{quote}
\textbf{Principle 8: Artefact Integration:} Workshop outputs connect to existing project artefacts and workflows (OKRs, PRDs, or others that the team is aiming for), translating ethical commitments into forms that can be tracked and enacted within normal project rhythms. This can connect workshop insight and ongoing practice, and help maintain attention to ethical considerations amid day-to-day operational pressures. \emph{Addresses:} Isolation from practice; Temporal discounting (partial)
\end{quote}

\begin{quote}
\textbf{Principle 9: Scenario Method Framing:} Role-play and scenario-based exercises must be framed appropriately for professional contexts. Facilitators should explain the rationale for perspective-taking methods, acknowledge that such exercises may feel unfamiliar, and position them as structured tools for surfacing considerations. \emph{Addresses:} Role-play resistance
\end{quote}

Each principle addresses specific failure modes as detailed above. Appendix A provides a practitioner-oriented diagnostic reference mapping failure modes to warning signs, principles, as well as potential mitigation strategies.

\section{Discussion}

\subsection{Contribution}

This paper makes three contributions to responsible innovation scholarship:

First, by introducing epistemic trust as an organizing concept for understanding intervention success and failure, we provide theoretical grounding for what practitioners often describe intuitively as ``buy-in'' or ``engagement.'' The five dimensions we identify offer a diagnostic framework for assessing intervention design. The mechanism model (Figure 1) articulates how intervention design choices shape epistemic trust, how trust enables engagement, and how failure modes break this pathway.

Second, our typology identifies 24 distinct failure modes: 20 organized by epistemic trust dimension, plus four cross-cutting failure modes involving interactions between dimensions. These cross-cutting failures, in particular, are challenging because they involve trade-offs between competing design objectives. For example, a facilitator cannot simply maximize Authority without attending to how expertise creates hierarchy that undermines Inclusivity. 

Third, our design principles aim to provide real world guidance. The Epistemic Trust Failure Mode Atlas in the Appendix (Table~\ref{tab:atlas}) shows how each principle can be operationalized in workshop design. This structure can provide a complete pathway from theory to practice.

\subsection{Alignment, Co-optation, and the Limits of Team-Level Interventions}

The dimension of Alignment illustrates the well-established worry that ethical interventions can be absorbed into existing organizational dynamics while leaving core technical trajectories largely untouched \citep{vanOudheusden2014, Genus2018}. This is directly relevant for our mechanism claim. Since epistemic trust is partly cultivated through alignment with engineering goals, if Alignment is interpreted as endorsement of existing objectives, epistemic trust can function as \emph{legitimization}.

\citet{Balmer2015, Balmer2016} analyze this risk through the lens of role dynamics in post-ELSI collaboration. They document how social scientists and ethicists embedded in technoscience settings are cast into characteristic roles such as ``representative of the public,'' ``foreteller,'' ``critic/trickster,'' or ``reflexivity-inducer'', each shaped by power asymmetries and requiring ongoing negotiation. The aspirational role of genuine ``co-producer of knowledge'' therefore remains difficult to achieve. When facilitators are positioned as foretellers or critics, organizations can host their presence without allowing critique to reshape objectives and the intervention thus becomes contained.

Such role dynamics also bear on our proposed epistemic trust framework. The Authority dimension does not just concern technical credibility, but also how facilitators are positioned within organizational power structures. Similarly, the Inclusivity dimension must account for whose voices are amplified or suppressed by role assignments. The co-optation risk identified above can be understood as a failure to move beyond ``foreteller'' or ``social lubricant'' roles toward genuine co-production. When facilitators are cast as external commentators whose role is to raise concerns that practitioners then address (or dismiss), ethical responsibility remains located in the intervention.

Several features of the Moral Imagination methodology are designed to mitigate these dynamics, reflected in our design principles.

First, the voluntary nature of participation (Principle 2: Non-Adversarial Framing) distinguishes the program from compliance-oriented review. Teams choose to invest time in reflection they perceive as valuable, establishing a relationship different from mandatory review. This does not fully eliminate role dynamics, but it can shift the initial framing from ``ethics imposed on us'' toward ``ethics with which we have chosen to engage.''

Second, Principle 5 (Deliberative Ownership) explicitly positions participants as the locus of ethical responsibility, which aims to resist the ``foreteller'' role in which facilitators pronounce on ethical matters while practitioners listen. Facilitators provide scaffolding and pose questions instead of delivering verdicts. This aims at cultivating practitioners' capacity for ethical reasoning in contrast to providing definitive answers. This principle extends to supporting teams in questioning project directions; that is, asking whether they are solving the right problem, for whom, and under what assumptions.

Third, Principle 6 (Benevolent Challenge) addresses the tension between maintaining critical distance and achieving trust. Facilitators must challenge assumptions without being cast as adversarial critics. The framing of challenge as collaborative inquiry and combined with explicit acknowledgment of the constraints practitioners face, aims to position facilitators as partners in reflection. The credibility-connection trap (Authority $\times$ Relevance) identifies the risk that facilitators who achieve credibility through technical fluency may lose the critical distance needed to surface novel considerations. Benevolent Challenge is designed to deal with this tension.

Fourth, Principle 4 (Structured Voice Distribution) aims to ensure that critical perspectives including from junior team members or cross-functional partners can surface even when senior engineers or dominant voices would prefer consensus. This principle resists the ``social lubricant'' role in which facilitators smooth over disagreement.

Despite these considerations, there are constraints that the methodology cannot fully resolve. Facilitators are employed by the same organization as participants, which can create accountability dynamics that differ from independent academic engagement. The voluntary uptake model means teams most resistant to ethical reflection may never engage. The translation of insights into Responsibility Objectives, while designed to ensure follow-through (Principle 8: Artefact Integration), also risks domesticating ethical commitments into manageable deliverables that fit existing project management priorities. And, lastly, even successful cultivation of team-level epistemic trust cannot address structural conflicts between commercial imperatives and social values that require governance mechanisms beyond the scope of any workshop intervention.

Our more modest claim is therefore that cultivating epistemic trust at the team level is \textit{necessary} but not \textit{sufficient} for responsible innovation. It addresses a genuine barrier, namely the tendency for practitioners to dismiss ethics interventions without claiming to resolve deeper structural tensions. The epistemic trust framework can serve both as a design tool for intervention development and as a reflexive diagnostic for assessing whether a given collaboration enables substantive critical engagement.

\subsection{Limitations}

Several limitations of our research should be acknowledged.

First, we have not conducted a formal empirical evaluation of workshop outcomes. Our account derives from systematic reflection on facilitation practice rather than from controlled comparison or independent outcome assessment. We cannot establish causal claims about the effects of specific design choices, nor can we rule out self-serving bias in facilitator judgments. The failure modes and design principles we present are practitioner hypotheses, not validated findings.

Second, participation in the moral imagination program is voluntary. This introduces selection bias. Our findings describe failure modes among teams already disposed toward ethical engagement. Teams most resistant to ethical reflection do not appear in our data. The failure modes should be understood as challenges that arise even under favorable conditions, not as a comprehensive account of barriers to ethics integration. Our findings may not generalize to mandatory interventions.

Third, our experience is drawn from a single large technology company with particular cultural characteristics and engineering practices. Engineering culture varies across companies, sectors, and national contexts; failure modes salient in our context may differ elsewhere. The five dimensions and nine principles should be understood as heuristic frameworks warranting adaptation. 

Fourth, we have limited evidence regarding long-term effects of the intervention. Debriefs provide some indication of immediate uptake, but we lack systematic data on whether Responsibility Objectives are enacted, whether team norms shift durably, or whether ethical considerations remain salient as projects evolve. Longitudinal research would be necessary to assess lasting impact.

Fifth, the intervention operates at the team level and cannot address structural organizational factors that shape the context within which teams operate. Even successful cultivation of epistemic trust may be insufficient when structural factors militate against ethical action.

And lastly, while we have mapped design principles to failure modes and operationalized them in workshop practices, we have not empirically validated that the principles actually mitigate the failure modes they claim to address. The mappings in Table~\ref{tab:atlas} in the Appendix represent our best judgment based on facilitation experience, but systematic testing is required to establish efficacy.

\subsection{Future Research Directions}

These limitations point toward several avenues for future research.

First, the five dimensions of epistemic trust could be operationalized as measurable constructs and tested through survey-based, experimental, or mixed-method designs; for example, examining whether higher perceived relevance, inclusivity, or facilitator authority predicts greater uptake of ethical considerations in subsequent decision-making. Second, the typology and design principles invite comparative testing across interventions and contexts. Cross-intervention work could assess whether STIR and other integration approaches exhibit similar failure modes, and whether specific modes cluster by format (embedded researcher versus workshop) or setting (academic versus corporate). Relatedly, controlled comparisons could evaluate the efficacy of particular principles (e.g., whether Structured Voice Distribution measurably changes participation dynamics, or whether Dual-Expertise Facilitation reduces Authority failures). Finally, longitudinal studies could test the mechanism model more directly by tracking persistence of effects over time and identifying how epistemic trust translates into engagement and downstream outcomes under varying organizational conditions.

Beyond formal research, we intend the framework to be practically usable. For practitioners designing responsible innovation interventions, the five dimensions can function as a diagnostic checklist for anticipating where epistemic trust is most fragile, while the Epistemic Trust Failure Mode Atlas (Table~\ref{tab:atlas}) indicates which design principles to prioritize given likely points of failure.

\section{Conclusion}

This paper has proposed epistemic trust as a mechanism that can govern whether bottom-up ethics interventions take hold in engineering contexts. Based on our practitioner work, we present a typology of failure modes that shows how trust breaks down across five dimensions, and nine design principles for cultivating it. While these remain practitioner hypotheses, their value is twofold: for researchers, epistemic trust names a gap the integration literature has left undertheorized and supplies constructs that can be operationalized and tested. And for practitioners, the typology and principles offer diagnostic guidance that travels beyond our own methodology.

Cultivating ethical deliberation in engineering contexts requires attention to whether practitioners regard ethics interventions as useful. Epistemic trust names this condition. Without it, even well-designed interventions will fail to achieve the engagement that responsible innovation requires.

\section*{Acknowledgements}

We thank Johnny S{\o}raker for feedback on a previous draft, and participants at the 2026 Ethics and Responsible Innovation Conference at TUHH Hamburg for helpful comments.

\section*{Author Contributions}

\begin{sloppypar}\noindent CRediT: Benjamin Lange: Conceptualization, Data curation, Formal analysis, Investigation, Methodology, Project administration, Supervision, Writing -- original draft, Writing -- review \& editing; Geoff Keeling: Methodology, Writing -- original draft, Writing -- review \& editing; Kyle Pedersen: Writing -- review \& editing; Carmen Heringer: Writing -- review \& editing; Susan B. Rubin: Writing -- review \& editing; Ben Zevenbergen: Methodology, Writing -- original draft, Writing -- review \& editing; Amanda McCroskery: Methodology, Writing -- original draft, Writing -- review \& editing.\end{sloppypar}

\section*{Disclosure Statement}

The authors of this manuscript are/were employees of Google and/or Google DeepMind when writing and conducting research for this research article. Geoff Keeling, Kyle Pedersen, and Amanda McCroskery are employed by Google and Google DeepMind and own Alphabet stock as part of their standard compensation packages. Carmen Heringer, Susan B. Rubin, and Ben Zevenbergen conducted this work while at Google. Benjamin Lange conducted part of the research as a Visiting Research Scientist at Google.

\section*{Funding}

No funding was received for this work.

\section*{Declaration of Generative AI Use}

In preparing this manuscript, the authors used Claude Opus 4.5 (Anthropic) and Gemini 3 Pro (Google) for copyedits and stylistic improvements for clarity and to assist with the formatting of tables. These tools were used because the manuscript synthesizes contributions from seven co-authors and required harmonization of style and presentation. All outputs were reviewed and revised by the authors, who take full responsibility for the content of the manuscript.

\appendix
\setcounter{table}{0}\renewcommand{\thetable}{A\arabic{table}}
\section{Appendix A: Epistemic Trust Failure Mode Atlas for Ethics and Responsible Innovation Interventions}

This atlas summarizes recurrent failure modes observed across ethics and responsible innovation (RI) interventions, organized to support diagnosis, comparative analysis, and intervention design. Each failure mode is defined, associated with typical triggers, linked to observable indicators, and paired with mitigation strategies and potential risks of overcorrection.

The atlas can be used as a coding schema, formative evaluation tool, or comparative framework across intervention contexts.

\footnotesize
\begin{longtable}{@{}p{1.6cm}p{2.2cm}p{3.2cm}p{0.8cm}p{3.8cm}@{}}
\caption{Epistemic Trust Failure Mode Atlas}
\label{tab:atlas} \\
\toprule
\textbf{Dimension} & \textbf{Failure Mode} & \textbf{Warning Signs} & \textbf{Prin.} & \textbf{Mitigation} \\
\midrule
\endfirsthead
\multicolumn{5}{c}{\tablename\ \thetable{} -- continued from previous page} \\
\toprule
\textbf{Dimension} & \textbf{Failure Mode} & \textbf{Warning Signs} & \textbf{Prin.} & \textbf{Mitigation} \\
\midrule
\endhead
\midrule
\multicolumn{5}{r}{Continued on next page} \\
\endfoot
\bottomrule
\endlastfoot
Relevance & External imposition & Defensive posture; ``Are you going to block us?'' & P2 & Explain bottom-up methodology; reframe as team's exploration \\[0.5em]
& Abstraction without connection & Blank stares; discussion stays at principle level & P1 & Ground concepts in project-specific examples \\[0.5em]
& Scenario irrelevance & ``This doesn't apply to us''; low engagement & P1 & Revise scenario to team's actual concerns \\[0.5em]
& Conversational drift & Discussion shifts to personal values; loses technical anchor & P1 & Redirect to technology implications \\[0.5em]
\midrule
Inclusivity & Hierarchy reproduction & Senior members dominate; juniors silent & P4 & Written responses before verbal; call on juniors by name \\[0.5em]
& Premature consensus & Rapid agreement; no dissent voiced & P4 & Explicitly invite disagreement; assign devil's advocate \\[0.5em]
& Role-play resistance & Eye-rolling; dismissive comments; breaking character & P9 & Explain rationale; frame as structured tool not theater \\[0.5em]
\midrule
Agency & Passive reception & Participants wait for prompts; don't generate questions & P5 & Reduce facilitator talking; position participants as experts \\[0.5em]
& Responsibility deflection & ``That's a policy decision''; ``Legal would need to weigh in'' & P5 & Redirect to team's sphere of influence \\[0.5em]
& Calibration failures & Disengagement (too simple) or confusion (too complex) & P1 & Calibrate complexity to audience; adjust in real-time \\[0.5em]
& Facilitation rigidity & ``This feels like a checklist''; visible frustration & P5 & Pause; ask what team actually needs \\[0.5em]
& Facilitation drift & Unfocused discussion; unclear objectives & P5 & Reassert structure; summarize and redirect \\[0.5em]
\midrule
Authority & Technical incredibility & Exchanged glances after misstatement; withdrawal & P3 & Acknowledge error; have technical co-facilitator step in \\[0.5em]
& Idiom dominance & Discussion becomes purely technical OR purely philosophical & P3 & Balance idioms; translate between domains \\[0.5em]
& Perceived na\"{i}vet\'{e} & ``You don't understand our constraints'' & P6 & Acknowledge constraints explicitly; demonstrate familiarity \\[0.5em]
& Challenge perceived as attack & Defensive responses; hostility to questions & P6 & Reframe as collaborative inquiry; model humility \\[0.5em]
\midrule
Alignment & Adversarial framing & Ethics perceived as constraint; ``ethics police'' & P2 & Position ethics as resource for achieving objectives well \\[0.5em]
& Temporal discounting & ``We don't have time for this'' & P8 & Connect to current sprint; frame as risk mitigation \\[0.5em]
& Goal divergence & Mismatch between expectations; ``not what we signed up for'' & P7 & Surface expectations at intake; renegotiate if needed \\[0.5em]
& Isolation from practice & Insights not enacted; no trace in project artefacts & P8 & Translate to native formats (OKRs, PRDs); schedule follow-up \\[0.5em]
\midrule
Cross-cutting & Credibility-connection trap & Technical facilitator accepts framings uncritically & P6 & Earn credibility, then exercise critical distance \\[0.5em]
& Participation paradox & Structured exercises feel forced; reluctant responses & P4+\allowbreak P5 & Enable voice without scripting; offer multiple modes \\[0.5em]
& Co-optation risk & Ethics instrumentalized; critical capacity neutralized & P5+\allowbreak P7 & Maintain space to challenge project direction \\[0.5em]
& Expertise hierarchy & Facilitator dominates; participants defer & P4+\allowbreak P6 & Emphasize questions over claims; invite disagreement \\
\end{longtable}

\end{document}